# Bone Material Analogues for PET/MRI Phantoms

Dharshan Chandramohan[1], Peng Cao[1,#], Misung Han[1], Hongyu An[2], John J. Sunderland[3], Paul E. Kinahan[4], Richard Laforest[2], Thomas A. Hope[1,*], Peder E. Z. Larson[1,*]

[1]Department of Radiology and Biomedical Imaging, University of California - San Francisco, San Francisco, CA, USA

[2]Department of Radiology, Washington University, St. Louis, MO, USA

[3]Departments of Radiology, Radiation Oncology, and Physics and Astronomy, University of Iowa, Iowa City, IA, USA,

[4]Department of Radiology, University of Washington, Seattle, WA, USA,

[#]Current Affiliation: Department of Diagnostic Radiology, The University of Hong Kong, Hong Kong, China

*These authors contributed equally

**Running Title**: PET/MRI Bone Phantoms

**Corresponding Author**:
Peder E. Z. Larson
1700 4th St
Byers Hall, Room 102C
San Francisco, CA 94158
peder.larson@ucsf.edu

**Abstract**

*Purpose*: To develop bone material analogues that can be used in construction of phantoms for simultaneous PET/MRI systems.

*Methods*: Plaster was used as the basis for the bone material analogues tested in this study. It was mixed with varying concentrations of an iodinated CT contrast, a gadolinium-based MR contrast agent, and copper sulfate to modulate the attenuation properties and MRI properties (T1 and T2*). Attenuation was measured with CT and 68Ge transmission scans, and MRI properties were measured with quantitative ultrashort echo time pulse sequences. A proof-of-concept skull was created by plaster casting.

*Results*: Undoped plaster has a 511 keV attenuation coefficient (~0.14 cm$^{-1}$) similar to cortical bone (0.10-0.15 cm$^{-1}$), but slightly longer T1 (~500 ms) and T2* (~1.2 ms) MR parameters compared to bone (T1 ~ 300 ms, T2* ~ 0.4 ms). Doping with the iodinated agent resulted in increased attenuation with minimal perturbation to the MR parameters. Doping with a gadolinium chelate greatly reduced T1 and T2*, resulting in extremely short T1 values when the target T2* values were reached, while the attenuation coefficient was unchanged. Doping with copper sulfate was more selective for T2* shortening and achieved comparable T1 and T2* values to bone (after 1 week of drying), while the attenuation coefficient was unchanged.

*Conclusions*: Plaster doped with copper sulfate is a promising bone material analogue for a PET/MRI phantom, mimicking the MR properties (T1 and T2*) and 511 keV attenuation coefficient of human cortical bone.

**Keywords:** PET/MRI, bone materials, attenuation correction, doped plaster

**Introduction**

Quantitative accuracy in PET/MRI studies is crucial for comparison between scans acquired at different times and on different equipment, and it is also an important factor in establishing benchmarks for clinical trials[1]. One of the major PET/MRI challenges for quantitative PET accuracy is to derive the 511 keV attenuation coefficient maps. In PET/CT systems, the CT provides a more direct measure of electron density that has been shown to correlate with PET attenuation[2], but MRI measures nuclear spin density and properties that do not directly correlate with PET attenuation. Creating MR-derived attenuation maps for PET/MRI is an active area of research[3–6].

Existing plastic/water-based phantoms do not provide a way of validating the MRI-derived attenuation map needed for quantitative PET imaging. A partial solution that has been used is to employ template attenuation maps (mu-maps) for standard calibration phantoms (e.g. NEMA IQ) to validate PET quantitation.These are however limited to a few geometries, require exact alignment with the PET data and do not account for potential errors in mu-maps derived from the MRI-based attenuation correction[7,8]. Furthermore, approaches for generating attenuation maps are varied and there is no industry consensus as to which method to use[9,10].

Of the different physiological tissues, bone offers a unique challenge because it has little to no signal in MRI due to its rapid relaxation rate and low proton density, yet it has the highest attenuation for PET photons[9]. Soft tissues are easier to characterize in MR-derived attenuation maps, as they have high signal on MRI and because the driver of differences in attenuation is fat versus water content, which is easy to measure with MRI. Furthermore, there are no prior bone phantom materials that satisfy both the PET and MRI properties required, whereas soft-tissue analogues with appropriate properties are easier to create from existing MRI phantoms[11] and with the requirement that the 511 keV attenuation coefficient equal that of water. Therefore we believe that in order to evaluate and validate PET/MRI attenuation correction methodologies in a controlled manner, appropriate bone-mineral analogues are needed in order to design anthropomorphically accurate body phantoms.

Most currently developed bone material analogues for imaging phantoms target either photon attenuation properties or MR properties but not both[12,13]. For phantoms mimicking the electron density and attenuation, $K_2PO_4$ and barium doped thermoplastics and polymer resins have been developed[14] and a density matched bone-mimicking material is incorporated in the American

College of Radiology (ACR) CT accreditation phantom. With respect to MR properties there have been very few studies. Prior work has identified a 3D printing material that was MR visible and printed bony anatomy, but had a low density[13], as well as a resin with MR properties that match cortical bone[12].

The purpose of this study is to investigate various potential formulations for bone mineral analogues using plaster as the baseline material and using precisely varying doping of the material to optimize MR relaxation properties and PET attenuation coefficient.

**Methods**

*Materials*

Plaster was used as the basis for the bone material analogues tested in this study. Plaster is a cement made by mixing calcium sulfate hemihydrate ($CaSO_4 \cdot 0.5H_2O$) with water to form the mineral gypsum ($CaSO_4 \cdot 2H_2O$). Plaster was chosen as a bone mineral analogue for three main reasons. First, plaster has historically been used as an orthopedic bone void filler due to the similarity in calcium content, mineral structure, and biomechanical properties of gypsum to hydroxyapatite. Second, plaster is facile to construct into a stable anthropomorphic geometry by filling a mold or by using a binder jetting 3D-printing process. Finally, it is easy to dope the water used in binding the plaster in order to manipulate properties of the material to more accurately resemble bone in terms of its MRI relaxation and PET attenuation properties.

The water used to form the plaster was doped with three materials, each targeting a separate imaging parameter. An iodinated CT contrast, Iohexol (Omnipaque, GE), was used to attempt to increase attenuation. A gadolinium-based MR contrast agent, Gadodiamide (Omniscan, GE), was used to shorten the MR $T_1$ and $T_2^*$ relaxation times, as was copper sulfate ($CuSO_4 \cdot nH_2O$). Two different ranges of concentrations were used for each of the three doping agents, resulting in six phantom assemblies. The first three phantoms constructed used the initial "exploratory" ranges, and the second set of phantoms constructed used more "targeted" ranges of concentrations intended to correspond to the range of physiological variation. Doping concentrations for all phantoms are listed in *(Table 1)*.

| Gd agent (% by mass) | | Cu agent (% by mass) | | I agent (% by mass) | |
|---|---|---|---|---|---|
| Exploratory | Targeted | Exploratory | Targeted | Exploratory | Targeted |
| 0.0%, | 0.0%, | 0.0%, | 2.0%, | 0.0%, | 0.0%, |
| 0.1%, | 0.02%, | 0.25%, | 2.4%, | 0.25%, | 2.4%, |
| 0.2%, | 0.04%, | 0.5%, | 2.8%, | 0.5%, | 4.8%, |
| 0.4%, | 0.06%, | 1.0%, | 3.2%, | 1.0%, | 7.2%, |
| 0.8%, | 0.08%, | 2.0%, | 3.6%, | 2.0%, | 9.6%, |
| 1.6% | 0.1% | 4.0% | 4.0% | 4.0% | 12.0% |

**Table 1.** Concentrations of the doping agents by mass in both the exploratory and target doping ranges

Phantoms were constructed by filling uniform 25 mL cylindrical high-density polyethylene screw-cap vials with the doped plaster mixtures. The vials were capped before curing of the plaster; care was taken to minimize the presence of air bubbles in the vials. Each phantom consisted of 19 containers arranged in a hexagonal close packing **(Fig. 1)**. Seven containers, including the center container, were filled with doped water solutions that ensured sufficient signal for successful MR scan calibration. The remaining twelve containers were filled with six different plaster concentrations; one of each pair of samples was placed nearer the center of the phantom and one was placed near the outside of the phantom to control for possible slow-varying intensity inhomogeneities in the images. MR scans of the "exploratory" phantom were acquired four times over the course of a month, and once three months after construction, while the 'targeted' phantom was scanned three times over the course of a month. $^{68}$Ge transmission scans were acquired once for each phantom after all MR scans had been acquired.

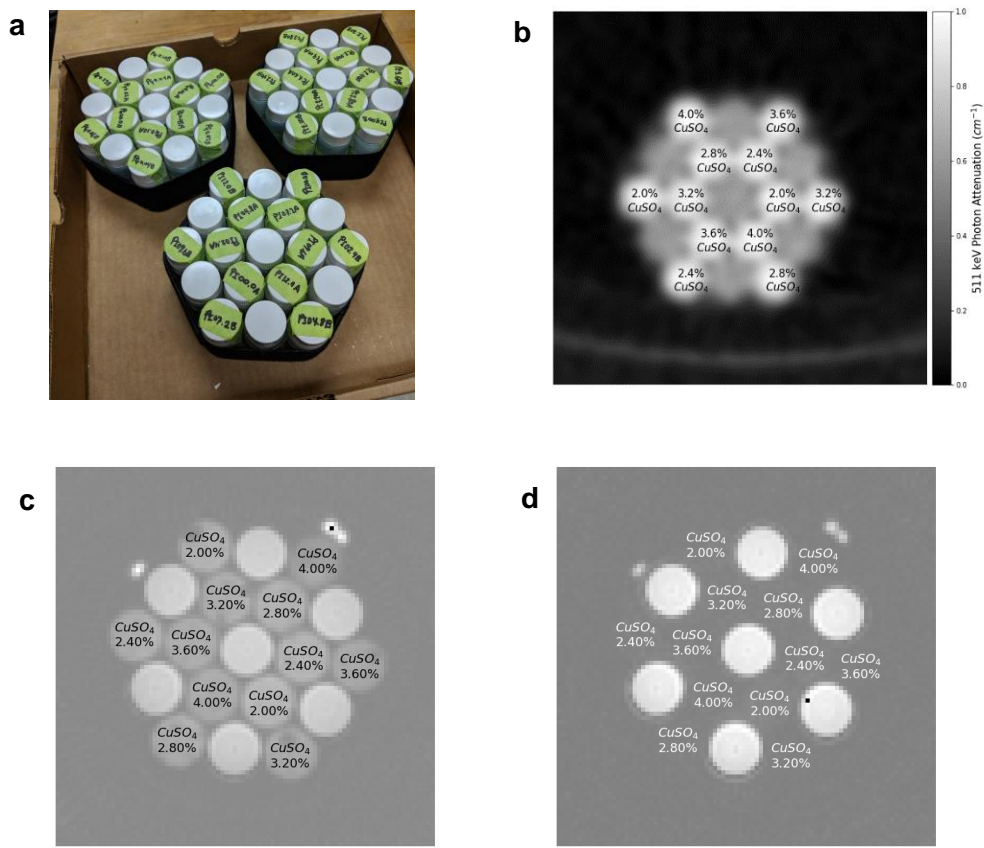

**Figure 1.** The testing phantom design consisted of polyethylene vials filled with doped or undoped plaster as well as vials filled with water for calibration. (a) Photograph of the vial assembly for testing. Examples of (b) $^{68}$Ge Transmission scan of copper sulfate-doped vials (c) UTE image at TE = 24 microseconds and (d) later TE image of copper sulfate-doped vials.

## *Casting plaster in anthropomorphic geometries*

As a proof of concept for an anthropomorphic phantom we used matrix casting to build phantoms from undoped and copper sulfate-doped plaster in the shape of the posterior half of a skull. We generated posterior skull molds from 3D-printed PLA plastic prototypes using silicone rubber and filled them with undoped plaster and plaster doped with 3% by weight copper sulfate, based on our initial results that the $T_2^*$ of plaster at this concentration of copper sulfate doping fell within the physiological range. Plaster doped with 3% copper sulfate proved difficult to pour to fill the mold as the plaster solidified much more quickly than undoped plaster, the resulting cast had a rough surface and inconsistent thickness.

*Imaging*

Quantitative MR parameter mapping was performed on a 3T GE MR750 scanner (GE Healthcare, Waukesha, WI). All scans for MR quantification used a 3D gradient echo sequence with ultrashort echo times (UTE) and radial readouts in order to maximize signal from the plaster **(Fig. 1)**. A single-channel quadrature head coil was used to provide more uniform B1 homogeneity. $T_1$ quantification used a variable flip angle (VFA) approach[15,16] with flip angles of 8 and 44 degrees; the TE for these scans was 24 microseconds, with a TR of 6.4 milliseconds. $T_2^*$ quantification was performed using a multi-echo sequence with 32 echo times ranging from 24 microseconds to 5 milliseconds[17]; these scans used an 8 degree flip angle and a TR of 9.276 milliseconds. The reconstructed images were 256x256 with 108 slices and 2mm isotropic voxels.

$^{68}$Ge transmission scans were acquired on a Siemens HR+ scanner. This scanner is equipped with 3 rotating 68Ge lines source (approximately 70MBq each), and 45 minutes transmission scans were performed for each arrangement of 19 containers.. These data provided direct quantitative measurement of the 511 keV photon attenuation coefficients of the different plaster formulations.

*Image processing and analysis*

Cylindrical regions of interest (ROIs) were placed in the center of each container with a radius half that of the container and a length half the container length as well. This was done to minimize the possibility of partial voluming from the container walls and possible air bubbles, and to minimize boundary effects. $T_1$ was calculated using the following equation (VFA method)

$T_1 = -TR / \ln((S_1/\sin(a_1) - S_2/\sin(a_2))/(S_1/\tan(a_1) - S_2/\tan(a_2)))$         (1)

where TR is the repetition time of the UTE sequence, $a_1$ and $a_2$ are the two flip angles used (8 and 4 degrees, respectively), and $S_1$ and $S_2$ are the corresponding signal magnitudes. An example of the implementation of this method for our data is illustrated in **(Fig. 2)**.

$T_2^*$ was calculated by pooling the voxels within the ROI and fitting the noise-corrected power of the complex UTE signal

$$P_{corr} = P_0 \exp(-2TE/T_2^*) \qquad (2)$$

Where $P_{corr} = S^2 - 2\sigma^2$ with $S^2$ being the square of the observed signal and $2\sigma^2$ being the noise estimate that can be computed as the mean power in an ROI devoid of signal. This power fitting avoids bias due to non-zero mean noise in the MRI magnitude[18]. Since we know that the plaster signal decay is rapid we used the mean signal in a plaster ROI on the last echo to estimate the noise. Fitting the model was performed using the SciPy implementation of the Levenberg-Marquardt algorithm for non-linear least squares estimation. An example of this fit for our data is illustrated in **(Fig. 2)**.

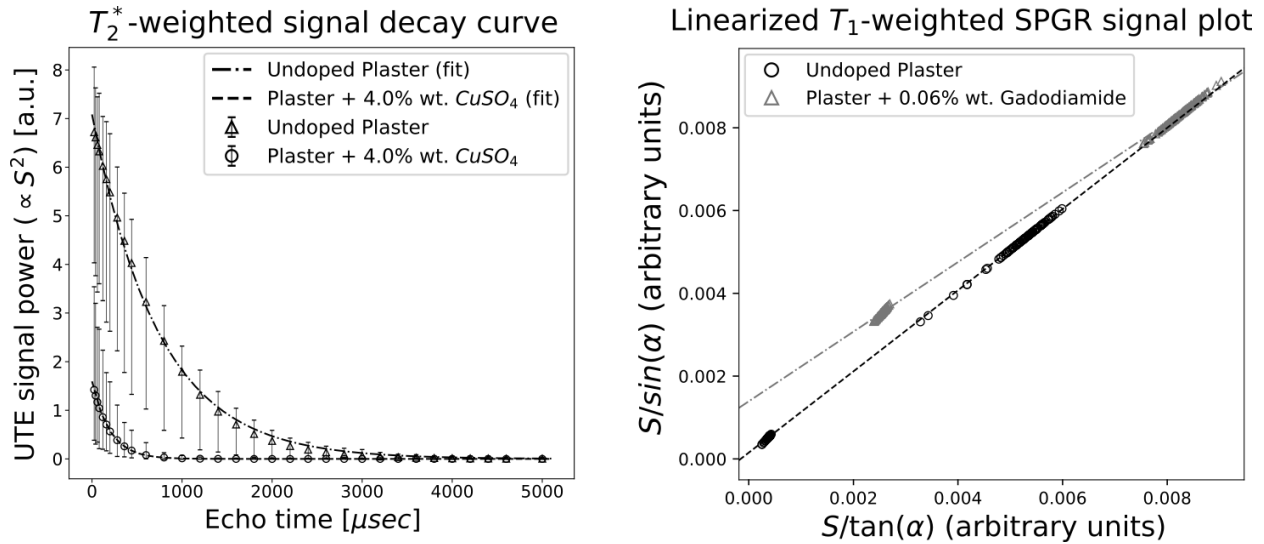

**Figure 2.** Sample data fitting. (left) T2* was quantified based on multiTE MRI and was fit to the noise-corrected power of the MR signal. (right) T1 was quantified using a variable flip angle method.

CT attenuation values measured in Hounsfield units were averaged over the matching ROI on the CT images. 511 keV photon attenuation values derived from the $^{68}$Ge transmission scans were averaged over an ROI of approximately the same size in approximately the same position.

**Results**

We targeted MR relaxation parameter values of $T_1$ between 200 and 250 ms[19–21] and $T_2^*$ between 300 and 500 us[19–22], as well as a 511 keV attenuation coefficient between 0.10 and 0.15 cm$^{-1}$. The results of the $T_2^*$, $T_1$, and attenuation coefficients are show in **Figures 3, 4, 5, and 6**, respectively, and the targeted parameter ranges are highlighted as well.

Undoped plaster had a measured $T_1$ of between 250 and 820 ms, measured $T_2^*$ of between 1070 and 1330 microseconds, and measured 511 keV attenuation coefficients between 0.140 cm$^{-1}$ and 0.148 cm$^{-1}$.

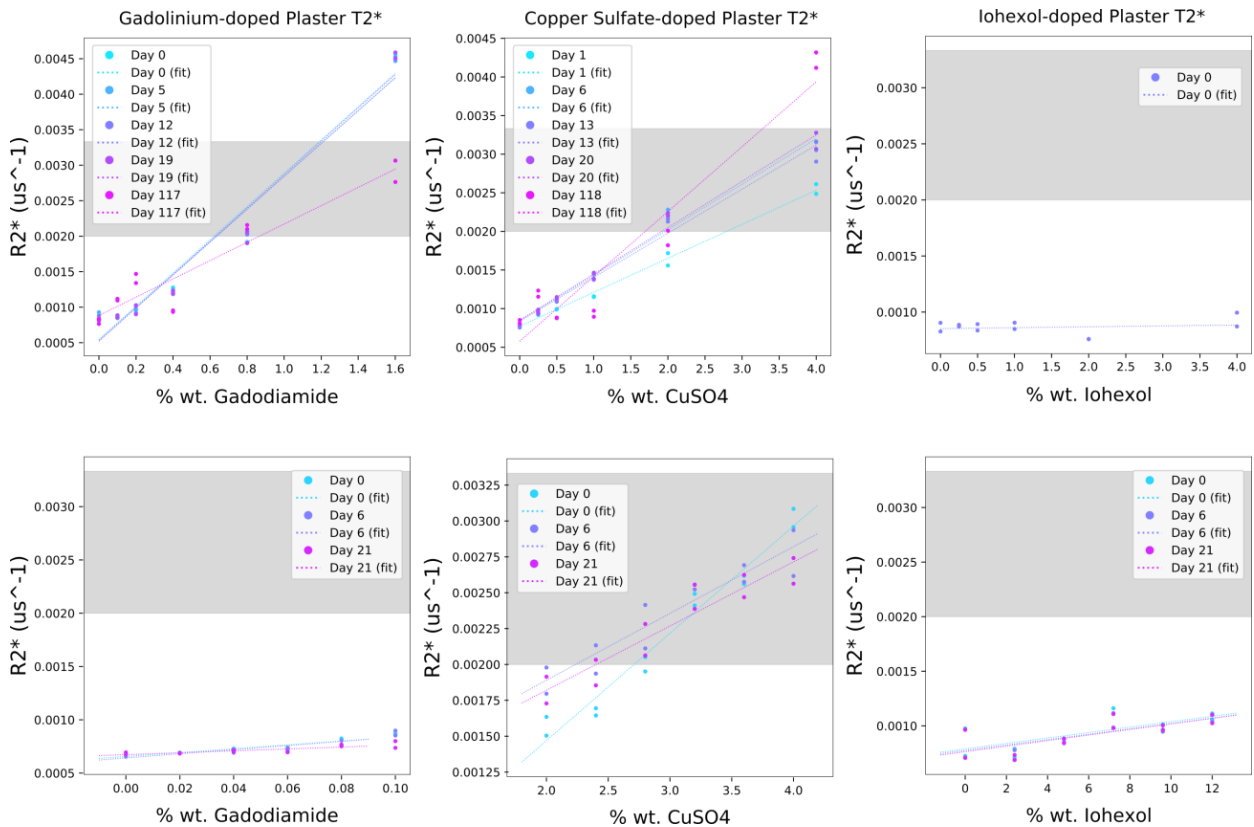

**Figure 3.** T2* results. The doped plaster R2* relaxation rates were measured within one day after mixing as well as several later times after preparation. Each point represents an individual vial (two vials were prepared for each concentration), and the dashed lines show linear fits to the R2* versus concentration. The gray shaded area indicates our targeted T2* range for dense bone of 300 to 500 microseconds which is based on reference [19-22]. The top row is from the Exploratory Concentrations, and the bottom row from the Targeted Concentrations.

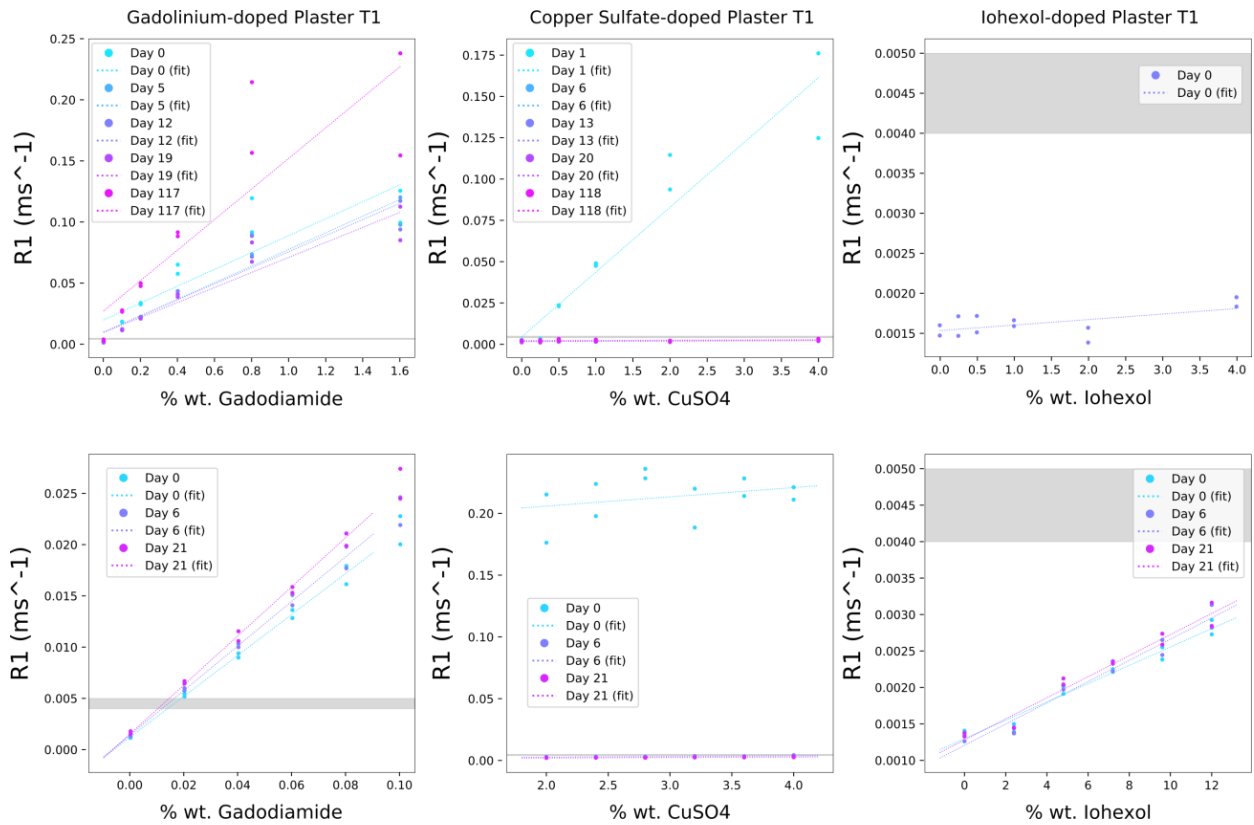

**Figure 4.** T1 results. The doped plaster R1 relaxation rates measured within one day after mixing as well as several later times after preparation. Each point represents an individual vial (two vials were prepared for each concentration), and the dashed lines show linear fits to the R1 versus concentration. The gray shaded area indicates our targeted T1 range for dense bone of 200 to 250 milliseconds which is based on reference [19-21]. The top row is from the Exploratory Concentrations, and the bottom row from the Targeted Concentrations.

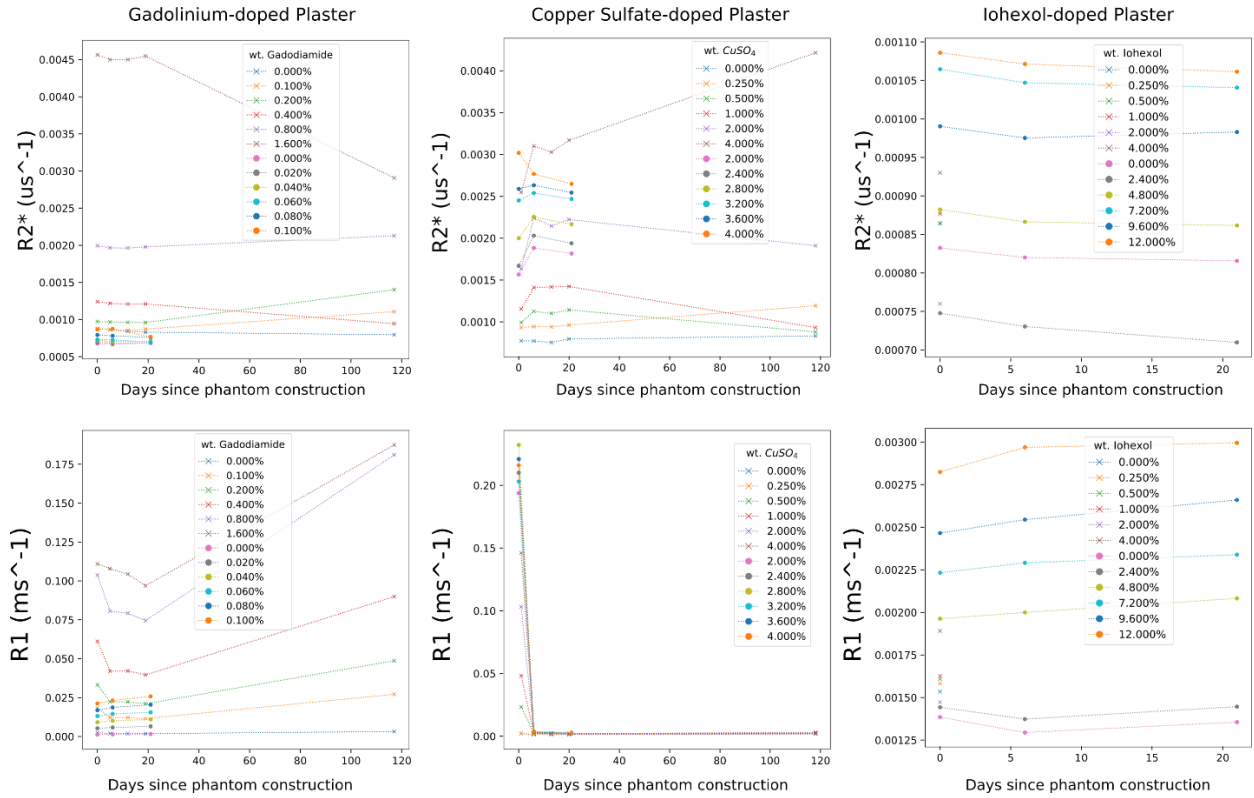

**Figure 5.** Changes in relaxation rate over time since phantom construction. This plot summarizes the temporal changes in relaxation rates for all plaster doping preparations. Each data point represents the average of all vials for a given concentration. This data was from the same experiments as **Figs. 3 and 4**, and exploratory ranges are plotted with x's and the targeted concentrations with circles.

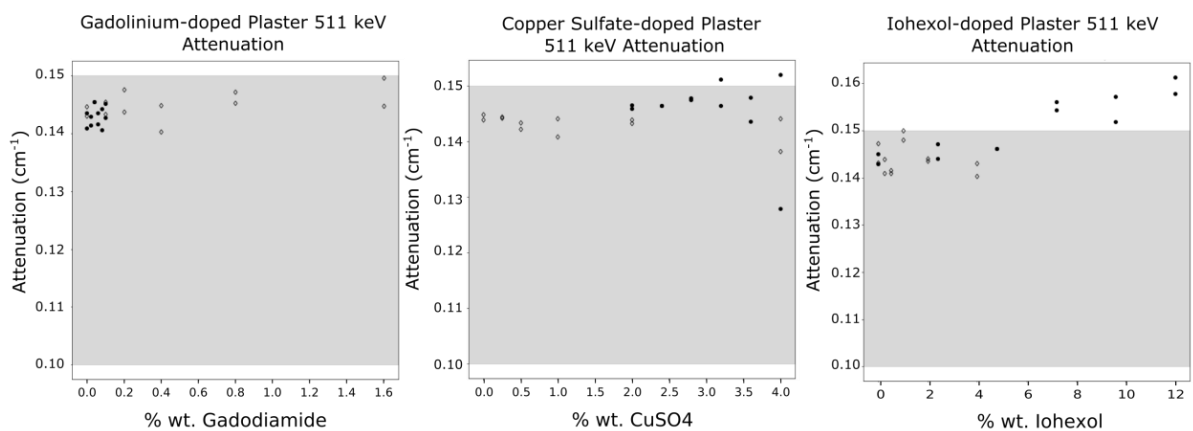

**Figure 6.** 511 keV photon attenuation results. The doped plaster samples 511 keV photon attenuation was measured a $^{68}$Ge transmission scan. The open circles show the values for the "exploratory" mixtures, while the closed circles are the values for the "target" mixtures. The gray shaded area indicates our targeted attenuation range for dense bone of 0.10 to 0.15 cm$^{-1}$.

*Exploratory phantoms*

Doping with gadolinium causes significant shortening of $T_1$. Doping plaster with gadodiamide causes $T_1$ to shorten to between 4 and 90 ms. The $T_1$ shortening increases with greater quantities of gadodiamide doping as we would expect. The $R_1$ relaxation rate was relatively stable for the scans up to Day 19 after mixture preparation. It was larger at the Day 117 scan, though the residuals are large for the linear fit, especially for containers with a large amount of gadodiamide doping.

Gadodiamide doped samples had a measured $T_2^*$ of between 210 microseconds and 1180 microseconds. $T_2^*$ shortening increases with greater quantities of gadodiamide doping. This relaxation rate appears fairly stable across measurements taken in the first month, with the final Day 117 measurement seeming to indicate a reduced $R_2^*$ effect.

Gadolinium doping across the broad range of concentrations in the exploratory phantom does not appear to significantly impact the attenuation of 511 keV photons in plaster.

Doping with copper sulfate causes a strong shortening of $T_1$ with increasing concentration of copper sulfate shortly after mixing. However, this prominent initial effect dissipates in measurements taken more than six days after the construction of the phantom. On the day the phantom was constructed $T_1$ values vary between 5 and 600 milliseconds, however, after one week copper sulfate-doped phantoms have a measured $T_1$ of between 300 and 800 milliseconds, with no obvious relationship with the copper sulfate concentration.

Copper sulfate doping significantly shortens $T_2^*$. Values for samples doped with copper sulfate range from 230 microseconds to 1140 microseconds. $T_2^*$ tends to decrease with increasing copper sulfate doping and the observed $R_2^*$ was relatively stable over time.

The amount of copper sulfate doping does not appear to significantly impact the attenuation of 511 keV photons in plaster over this broad range of concentrations.

Iohexol doping in this exploratory range of concentrations does not appear to have any significant effect on $T_1$ or $T_2^*$ relaxation times. Values of $T_1$ and $T_2^*$ do not differ significantly between doped and undoped samples. Across this range of quantities of iohexol doping the 511 keV photon attenuation values also appear to cluster around the value for undoped plaster.

*Targeted concentration phantoms*

Gadodiamide doping still has a strong effect on $T_1$ shortening across the physiologically targeted range of concentrations . Gadodiamide-doped samples had $T_1$ values ranging from 30 to 200 milliseconds. $T_1$ shortening increases linearly even across this narrow range of concentrations. The relaxation rate is moderately stable over the course of a month, though the effect increases slightly with time.

At these small concentrations the effect of gadodiamide doping on $T_2^*$ is negligible. $T_2^*$ values for doped samples range from 1100 to 1460 microseconds, with a slight shortening of $T_2^*$ with increasing gadodiamide concentration. This relatively small effect is consistent across all time points.

Attenuation values for 511 keV photons over this range of gadodiamide quantities clusters very closely around the value for undoped plaster.

Copper sulfate doping between 2% and 4% by mass causes significant $T_1$ shortening immediately after the sample is mixed, though the effect goes away when measured one week later. The initial $T_1$ values were measured between 4 and 6 milliseconds, however after one week $T_1$ values are measured between 220 and 480 milliseconds, which is in the same range of values for undoped plaster. Beyond 1 week, the $T_1$ values were relatively stable.

There is a fairly strong reliable linear effect of copper sulfate doping on $T_2^*$ within this targeted range of concentrations. $T_2^*$ values range between 320 and 620 microseconds. The relaxation rate is fairly consistent between the measurements taken one week and three weeks after the samples were mixed, while initially the $T_2^*$ appeared slightly elevated.

511 keV photon attenuation at these concentrations of copper sulfate doping appear slightly elevated relative to the attenuation coefficient of undoped plaster. Copper sulfate-doped samples have attenuation coefficients ranging between 0.145 cm$^{-1}$ and 0.152 cm$^{-1}$, while undoped plaster attenuation coefficients range between 0.140 cm$^{-1}$ and 0.148 cm$^{-1}$

With a high enough concentration of iohexol doping we were able to observe a shortening of $T_1$. Iohexol-doped samples had $T_1$ values ranging from 340 to 720 milliseconds with shorter values observed for higher quantities of doping. The linear effect was preserved consistently across all time points.

At these high iohexol concentrations a slight shortening of $T_2^*$ was also observed. Measured $T_2^*$ values ranged between 900 and 1460 microseconds. These values are still fairly close to the $T_2^*$ value for undoped plaster, but there is a slight linear effect of iohexol concentration on $T_2^*$ that remains consistent across the three time points surveyed.

Iohexol doping does appear to significantly increase the attenuation of doped plaster to 511 keV photons. Attenuation coefficients of doped samples range from 0.144 to 0.162 cm$^{-1}$.

### *Relaxation Rates over Time*
Changes in the relaxation rates in the time since phantom construction is shown in **Fig. 5**. The most notable changes in relaxation rate were in the copper sulfate $T_1$ from within one day of construction to day 6 after construction (statistically significant, $p < .05$, paired T-test). All other temporal changes in relaxation rate were of a much smaller magnitude. There was a significant change in the gadodiamide $T_1$ from Day 19 to Day 117 ($p < .05$).

### *Casting plaster in anthropomorphic geometries*

In a first casting experiment, we created and compared a first set of doped and undoped casts scanned with the $T_2^*$ quantification protocol (**Fig. 7**). We observed a shortening of $T_2^*$ for the doped plaster cast relative to the undoped cast, however, the extent of the effect was not what we had observed in containers. The casts were scanned within a week of construction. After the casts had initially set they were left in open air, potentially leading to different drying characteristics than that of plaster sealed in small vials.

In a second experiment, we created a second undoped plaster cast and also imaged the PLA plastic prototype. In the plaster we observed an increase in microbubbles on high resolution CT. We also observed $T_2^*$ values shorter than undoped plaster in vials and our first undoped plaster cast that were actually within the physiologic range for bone for this plaster phantom. We speculate the increased microbubbles shortened the $T_2^*$ due to local magnetic susceptibility differences.

The $T_2^*$ of PLA plastic was also quantified in this experiment, with values close to those for undoped plaster in previous experiments. However, the CT attenuation for PLA plastic (~ 300 HU, **Fig. 7**) is significantly less than that of plaster (~ 750 HU, **Fig. 7**) and of bone (500-2000 HU).

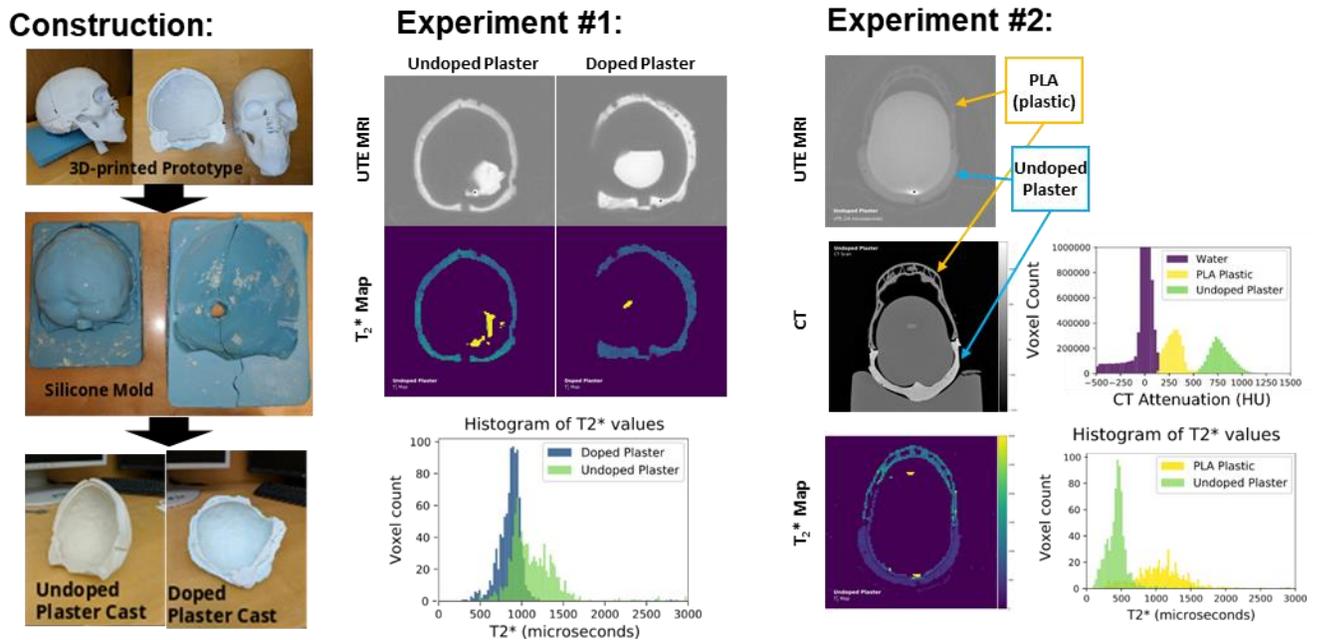

**Figure 7.** Proof of concept anthropomorphic skull cast using plaster. A 3D-printed PLA skull was used to create a silicone mold from which a plaster cast was created. In Experiment #1, an undoped and a copper sulfate doped (3.0 %wt $CuSO_4$) plaster cast of the posterior

skull half were compared.  In Experiment #2, an undoped plaster cast of the posterior skull half and the original PLA printed anterior skull half were imaged.

**Discussion**

*Comparing doped material properties to physiological values for bone*

Previous studies have measured physiological values for cortical bone of $T_1$ between 200 and 250 milliseconds, $T_2^*$ between 300 and 500 microseconds, and 511 keV attenuation coefficients between 0.10 cm$^{-1}$ and 0.15 cm$^{-1}$. Based on our measurements of undoped plaster, we targeted doping to decrease $T_1$ and $T_2^*$ while maintaining the attenuation coefficient.

Gadolinium exhibits such a strong $T_1$ shortening effect that it effectively overshoots the target $T_1$ values taken from the literature for bone at both high and low concentrations, however, given that the relaxivity is linear it is probable that smaller concentrations, less than 0.02% by mass, would fall within the range of physiological $T_1$ values for bone (between 200 and 250 milliseconds). At the concentrations required to fit this range gadodiamide would not be expected to significantly impact the $T_2^*$ or attenuation coefficient of the material.

Copper sulfate, on the other hand, has a strong effect on $T_2^*$, and after one week of drying it had almost no effect on $T_1$. The most likely explanation for this is that as the plaster dries the copper sulfate is in less proximity to the bound water, mitigating $T_1$ effects, while the $T_2^*$ effect, which is likely related to increased magnetic susceptibility, is still active over the larger distances between Cu ions and bound water molecules in the hydrated crystal. Doping values of between 2.8% and 4% by mass give $T_2^*$ values that fall within the physiological range of $T_2^*$ values (between 300 and 500 microseconds) with observed stability over time. Stability could potentially be increased by sealing the plaster with a waterproof coating, like polyurethane rubber, at some specified point after the material has been mixed, to prevent further loss of water once the MR relaxation properties are within the desired range.

Concentrations of copper sulfate above 3% by mass may have the unwanted effect of increasing the attenuation coefficient for 511 keV photons beyond the physiological range. Typical attenuation coefficients for bone range between 0.10 cm$^{-1}$ and 0.15 cm$^{-1}$. Because plaster is significantly dense on its own, undoped plaster is already at the high part of this range. Doping with excessive concentrations of copper sulfate could cause the attenuation coefficient to marginally exceed physiologically typical values.

Iohexol can be used to increase the attenuation coefficient if needed without significantly impacting the MR relaxation properties of the material. However, iohexol concentrations in excess of approximately 7% by mass might also cause an excessive increase in the attenuation coefficient that would exceed the physiological range.

Based on our results, doped plaster can be directly used for simulating dense bone based on the measured attenuation coefficient for 511 keV photons. Alternate mixing formulations may allow for simulating less dense bone. For example, mixing doped plaster with oil could provide a good simulation of trabecular bone and bone marrow, or mixing such that more air is trapped in the plaster could simulate less dense bone. Variations in mixing may also allow for undoped plaster to be used as a dense bone mimic, as undoped cast plaster in one case showed relaxation rates similar to dense bone (Experiment 2 in **Fig. 7**).

Our results have begun to demonstrate the long-term temporal stability, up to 118 days, of the MR and attenuation properties of doped plaster, but more investigation is warranted to create stable phantoms for calibration and qualification. We suspect much of the stability is related to the drying or dehydration of the plaster, with by far the most significant effect on $T_1$ with copper sulfate doping. Since we primarily used sealed vials, this probably stopped the drying process, but with plaster continually exposed to air more dehydration could occur. One possible solution would be to seal the plaster to lock in a certain water content.

3D-printed materials for finite deposition modeling printers tend to be thermoplastics with low photon attenuation, and often low MR signal if any. This is similar for resins used in stereolithography 3D printing. Plaster can be used as the material for some binder jetting process-based 3D printers, though such printers and plaster-based materials are rare and tend to be proprietary. Furthermore the binder jetting process for producing 3D printed plaster tends to reduce the density of the final part relative to cast parts. Casting undoped plaster in a mold can be an effective way to produce a phantom with an anthropomorphic geometry with accurate MR and PET attenuation properties.

**Conclusion**

Plaster is a promising bone material analogue for a PET/MRI phantom. In this paper, we demonstrated modulation of the MR properties (T1 and T2*) by doping plaster withcopper sulfate and gadodiamide, allowing for adjustment of these parameters to match in vivo MR properties. Without doping the 511 keV attenuation coefficient of plaster is similar to human cortical bone. Gadodiamide doping of plaster caused $T_2^*$ shortening at high concentrations, with significant $T_1$ shortening even at low concentrations. Copper sulfate doping of plaster caused significant $T_2^*$ shortening at moderate to high concentrations. Copper sulfate doping initially causes $T_1$ shortening but the effect was gone after about 1 week, presumably as the material dries. Iohexol doping of plaster has some effect on MR relaxation, causing minor shortening of both $T_1$ and $T_2^*$ at high concentrations, although the effect is potentially small enough to be negligible. Iohexol, however, does increase attenuation of the material for 511 keV photons. We also showed in a proof-of-concept experiment that undoped and doped plaster can be cast into an anthropomorphic shape while maintaining the attenuation and MR properties. The remaining challenges include refinement of anthropomorphic phantoms and better understanding of the plaster and doped plaster stability. In summary, plaster with these doping agents is very promising to create a bone-analogous material that is quantitatively bone-like for both MR relaxation and PET attenuation properties so could be used to evaluate PET/MRI attenuation correction methods.

**Acknowledgements**

This work was funded by the NIH/NCI grant # R01 CA212148. We would like to thank Karl Stupic, Kathryn Keenan, and Stephen Russek from the US National Institute of Standards and Technology (NIST) for helpful discussions and inspiration.

**Conflict of Interest Statement**

Dr. Larson and Dr. Hope received research funding support from GE Healthcare.

**References**

1. Ehman EC, Johnson GB, Villanueva-Meyer JE, et al. PET/MRI: Where might it replace PET/CT? *J Magn Reson Imaging*. March 2017:n/a-n/a. doi:10.1002/jmri.25711
2. Kinahan PE, Townsend DW, Beyer T, Sashin D. Attenuation correction for a combined 3D PET/CT scanner. *Med Phys*. 1998;25(10):2046-2053. doi:10.1118/1.598392
3. Paulus DH, Quick HH, Geppert C, et al. Whole-Body PET/MR Imaging: Quantitative Evaluation of a Novel Model-Based MR Attenuation Correction Method Including Bone. *J Nucl Med*. 2015;56(7):1061-1066. doi:10.2967/jnumed.115.156000

4. Liu F, Jang H, Kijowski R, Bradshaw T, McMillan AB. Deep Learning MR Imaging–based Attenuation Correction for PET/MR Imaging. *Radiology*. September 2017:170700. doi:10.1148/radiol.2017170700

5. Leynes AP, Yang J, Wiesinger F, et al. Direct PseudoCT Generation for Pelvis PET/MRI Attenuation Correction using Deep Convolutional Neural Networks with Multi-parametric MRI: Zero Echo-time and Dixon Deep pseudoCT (ZeDD-CT). *J Nucl Med*. October 2017:jnumed.117.198051. doi:10.2967/jnumed.117.198051

6. Torrado-Carvajal A, Vera-Olmos J, Izquierdo-Garcia D, et al. Dixon-VIBE Deep Learning (DIVIDE) Pseudo-CT Synthesis for Pelvis PET/MR Attenuation Correction. *J Nucl Med Off Publ Soc Nucl Med*. 2019;60(3):429-435. doi:10.2967/jnumed.118.209288

7. Ziegler S, Jakoby BW, Braun H, Paulus DH, Quick HH. NEMA image quality phantom measurements and attenuation correction in integrated PET/MR hybrid imaging. *EJNMMI Phys*. 2015;2. doi:10.1186/s40658-015-0122-3

8. Boellaard R, Rausch I, Beyer T, et al. Quality control for quantitative multicenter whole-body PET/MR studies: A NEMA image quality phantom study with three current PET/MR systems. *Med Phys*. 2015;42(10):5961-5969. doi:10.1118/1.4930962

9. Ladefoged CN, Law I, Anazodo U, et al. A multi-centre evaluation of eleven clinically feasible brain PET/MRI attenuation correction techniques using a large cohort of patients. *NeuroImage*. 2017;147:346-359. doi:10.1016/j.neuroimage.2016.12.010

10. Catana C, Quick HH, Zaidi H. Current commercial techniques for MRI-guided attenuation correction are insufficient and will limit the wider acceptance of PET/MRI technology in the clinic. *Med Phys*. 2018;45(9):4007-4010. doi:10.1002/mp.12963

11. Keenan KE, Ainslie M, Barker AJ, et al. Quantitative magnetic resonance imaging phantoms: A review and the need for a system phantom. *Magn Reson Med*. 2018;79(1):48-61. doi:10.1002/mrm.26982

12. Rai R, Manton D, Jameson MG, et al. 3D printed phantoms mimicking cortical bone for the assessment of ultrashort echo time magnetic resonance imaging. *Med Phys*. 2018;45(2):758-766. doi:10.1002/mp.12727

13. Mitsouras D, Lee TC, Liacouras P, et al. Three-dimensional printing of MRI-visible phantoms and MR image-guided therapy simulation. *Magn Reson Med*. 2017;77(2):613-622. doi:10.1002/mrm.26136

14. Hamedani BA, Melvin A, Vaheesan K, Gadani S, Pereira K, Hall AF. Three-dimensional printing CT-derived objects with controllable radiopacity. *J Appl Clin Med Phys*. 2018;19(2):317-328. doi:10.1002/acm2.12278

15. Deoni SCL, Rutt BK, Peters TM. Rapid combined T1 and T2 mapping using gradient recalled acquisition in the steady state. *Magn Reson Med*. 2003;49(3):515-526. doi:10.1002/mrm.10407
16. Han M, Rieke V, Scott SJ, et al. Quantifying temperature-dependent T1 changes in cortical bone using ultrashort echo-time MRI. *Magn Reson Med*. September 2015. doi:10.1002/mrm.25994
17. Boucneau T, Cao P, Tang S, et al. In vivo characterization of brain ultrashort-T2 components. *Magn Reson Med*. 2018;80(2):726-735. doi:10.1002/mrm.27037
18. Miller AJ, Joseph PM. The use of power images to perform quantitative analysis on low SNR MR images. *Magn Reson Imaging*. 1993;11(7):1051-1056. doi:10.1016/0730-725X(93)90225-3
19. Du J, Bydder GM. Qualitative and quantitative ultrashort-TE MRI of cortical bone. *NMR Biomed*. December 2012. doi:10.1002/nbm.2906
20. Reichert ILH, Robson MD, Gatehouse PD, et al. Magnetic resonance imaging of cortical bone with ultrashort TE pulse sequences. *Magn Reson Imaging*. 2005;23(5):611-618.
21. Techawiboonwong A, Song HK, Leonard MB, Wehrli FW. Cortical bone water: In vivo quantification with ultrashort echo-time MR imaging. *Radiology*. 2008;248(3):824–33.
22. Horch RA, Nyman JS, Gochberg DF, Dortch RD, Does MD. Characterization of 1H NMR signal in human cortical bone for magnetic resonance imaging. *Magn Reson Med*. 2010;64(3):680-687. doi:10.1002/mrm.22459